\documentclass[letterpaper, 10 pt, conference]{ieeeconf}  

\IEEEoverridecommandlockouts                              

\let\labelindent\relax

\usepackage{amsmath} 
\usepackage{amssymb}  

\usepackage{multicol}
\usepackage[bookmarks=true]{hyperref}
\usepackage{float}
\usepackage{graphicx}
\usepackage{amsfonts}
\usepackage{booktabs}
\usepackage{bbm}
\usepackage{graphics, wrapfig}
\usepackage{caption}
\usepackage{subcaption}
\usepackage{enumitem}
\usepackage{stackengine}
\usepackage{tikz}
\usepackage{dsfont}

\DeclareGraphicsExtensions{.png,.pdf,.eps,.jpg}

\newcommand{\calS}{{\mathcal S}}
\newcommand{\calA}{{\mathcal A}}
\newcommand{\calT}{{\mathcal T}}
\newcommand{\calR}{{\mathcal R}}
\newcommand{\calH}{{\mathcal H}}
\newcommand{\calD}{{\mathcal D}}
\newcommand{\calL}{{\mathcal L}}

\newcommand{\reftab}[1]{Table~\ref{tab:#1}}

\definecolor{darkred}{rgb}{0.6,0,0}
\definecolor{green}{rgb}{0.0,0.5,0}
\definecolor{blue}{rgb}{0,0,0.75}
\definecolor{orange}{rgb}{1,0.6,0.2}
\definecolor{red}{rgb}{1,0,0}

\newcommand{\citeneeded}[1]{\textcolor{red}{(citeme)}}

\title{\LARGE \bf
Learning Latent Traits for Simulated Cooperative Driving Tasks
}

\author{Jonathan A. DeCastro$^{1}$, Deepak Gopinath$^{1}$, Guy Rosman$^{1}$, Emily Sumner$^{2}$, Shabnam Hakimi$^{2}$, Simon Stent$^{1}$ 
\thanks{Toyota Research Institute,
        $^{1}$One Kendall Square, Cambridge, MA 02139, $^{2}$4440 El Camino Real, Los Altos, CA 94022 
        {\tt\small <firstname.lastname>@tri.global}}%
}

\begin{document}

\maketitle
\thispagestyle{empty}
\pagestyle{empty}

\begin{abstract}
To construct effective teaming strategies between humans and AI systems in complex, risky situations requires an understanding of individual preferences and behaviors of humans.  Previously this problem has been treated in case-specific or data-agnostic ways. In this paper, we build a framework capable of capturing a compact latent representation of the human in terms of their behavior and preferences based on data from a simulated population of drivers.  Our framework leverages, to the extent available, knowledge of individual preferences and types from samples within the population to deploy interaction policies appropriate for specific drivers.  We then build a lightweight simulation environment, \texttt{HMIway-env}, for modelling one form of distracted driving behavior, and use it to generate data for different driver types and train intervention policies.  We finally use this environment to quantify both the ability to discriminate drivers and the effectiveness of intervention policies.
\end{abstract}


\section{INTRODUCTION}
For robots to successfully cooperate with humans to solve complex tasks, it is important to equip them with a rich understanding of human behavior. 
For example, when solving a complex task such as driving, human drivers often vary in their preferences for when and how to receive help.
If a robot offers help at the wrong time or in the wrong manner, it can lead to a drop in task performance, reducing trust, and making further cooperation difficult.

An individual driver's receptivity towards help is a function of many things, such as the specifics of the task, the current situation, and the individual's own traits (such as tendency to be inattentive or make impulsive decisions) and preferences (such as willingness to accept help or tolerate risk).
Modern driver-assistance systems (e.g. lane-keep assist, forward collision warning systems) are \textit{not adaptive}, and intervene
irrespective of whether the driver is a teenage novice requiring momentary, repeated help to teach them to become a better driver, or an experienced driver suffering impairment and in need of more significant assistance.
This can lead to sub-optimal outcomes.
Developing more human-aware solutions to cooperation is challenging, but will allow assistance systems to be better tailored towards individuals, similar to how good driving instructors learn to adapt their styles to assist drivers with different needs.


\begin{figure}
    \centering
    \includegraphics[width=1.0\linewidth]{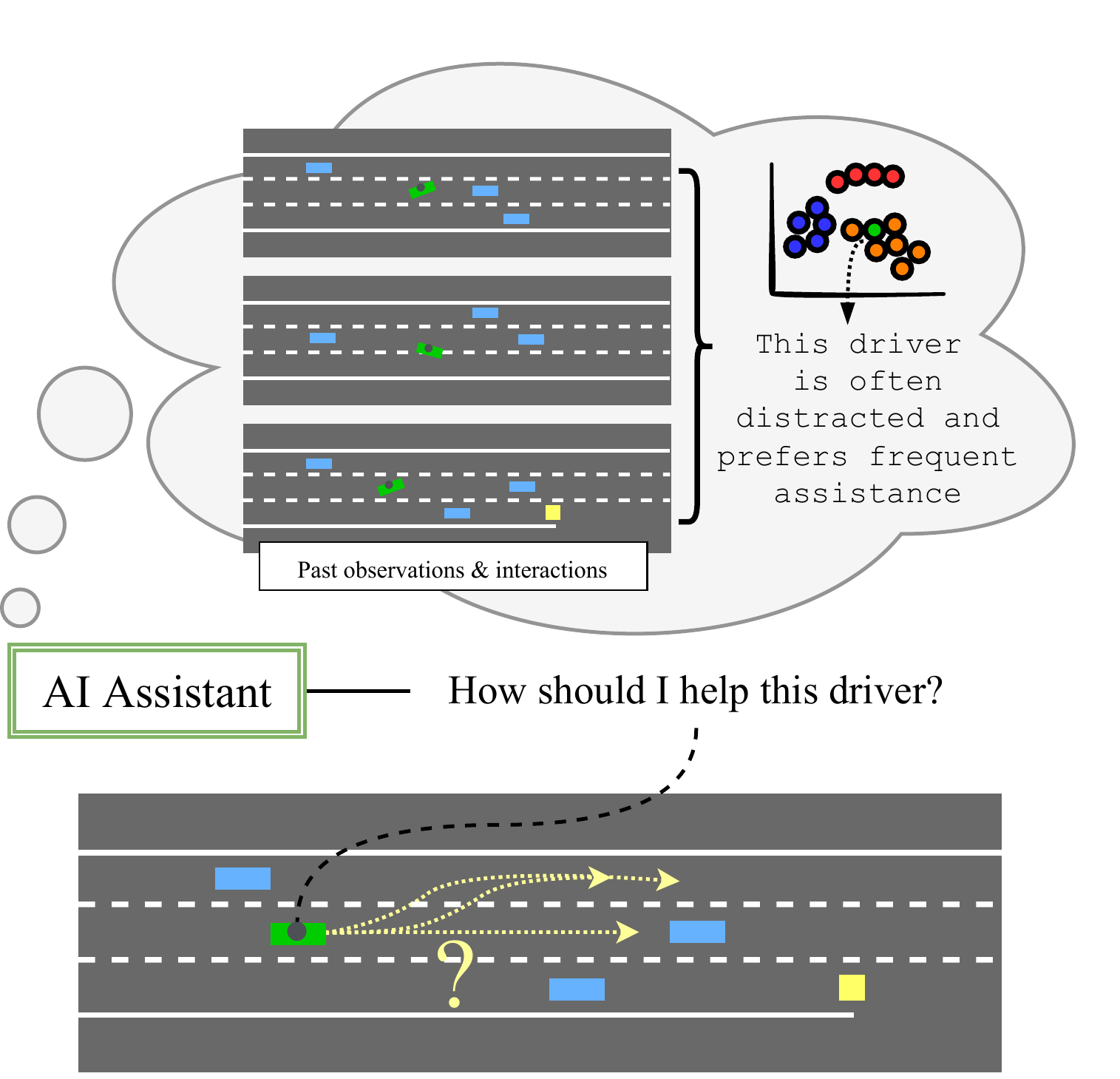}
    \caption{\textbf{Schematic of our approach.} In a cooperative driving task, we encode driver behavior characteristics, including preferences towards receiving assistance, within an embedding space, using knowledge of past observations and interactions. These preferences vary depending on the individual in a somewhat predictable manner, as certain agents may want to explore on their own without robot assistance while others may enjoy more frequent interactions.} 
    \label{fig:teaser}
\end{figure}

In this work, we posit that better solutions to cooperative human-machine teaming in driving can be achieved by first observing a given driver's behaviors over time, then using a summary of that behavior (in the form of a compact learned representation) to understand how to help that driver.  Rather than being reactive to the situation, our approach finds the best intervention strategy \textit{for the driver} by observing their driving behaviors in the setting of a population of drivers, for whom interactions policies have been designed.
Large-scale datasets are abundant in driving; however, due to practical considerations (privacy, cost, lack of human-centric measurements), they often do not include any direct measures of driver traits or preferences for receiving assistance.  Moreover, the majority of typical driving trips do not reveal differences between individuals.  Given these limitations, finding a representation that is useful for a machine to interact with the driver is a challenging problem.

We propose to tackle this question by embedding drivers in a latent space of possible behavior types, and using this embedding to specify policies and rewards that encode the human's intent.
A consequence of this is that we can leverage the policies and rewards to form cooperative teams~\cite{schaff2020residual, backman2022reinforcement} and the latent representations to deploy the best assistance modality for a given user (e.g. the type or frequency of intervention).  Latent representations have shown promise for personalization in domains outside of driving, such as sentiment analysis~\cite{Tanjim2020}, product recommendation~\cite{ijcai2017-382, Yu2019-vv}, intent recognition~\cite{Xie2020-dc}, and dialogue understanding~\cite{Yang2021-rj}.
We show, via simulation, that the approach holds promise in identifying traits that are important for aiding individual drivers at runtime, e.g. by helping to curb aggressive driving or to reduce distraction.  Figure~\ref{fig:teaser} illustrates the approach.  


\textbf{Key contributions.}
\begin{enumerate}
    \item We propose a framework for uncovering latent structure representing the traits of a driver based on patterns of driving observed over multiple demonstrations of behavior, and we tie this latent strucuture to the correct system for intervention.
    \item We introduce a new, simulation environment---\texttt{HMIway-env}---that captures different facets of human driving behavior; in particular, we use it to model distraction and cautiousness of human drivers, validated via human subjects.
    \item We use this environment as a basis for learning an assistive policy for interacting with a driver.
    \item We study the efficacy of our approach for driver inference with respect to different modeled cognitive factors in a simulated highway driving scenario.
\end{enumerate}

\begin{figure*}
    \centering
    \includegraphics[width=1\linewidth]{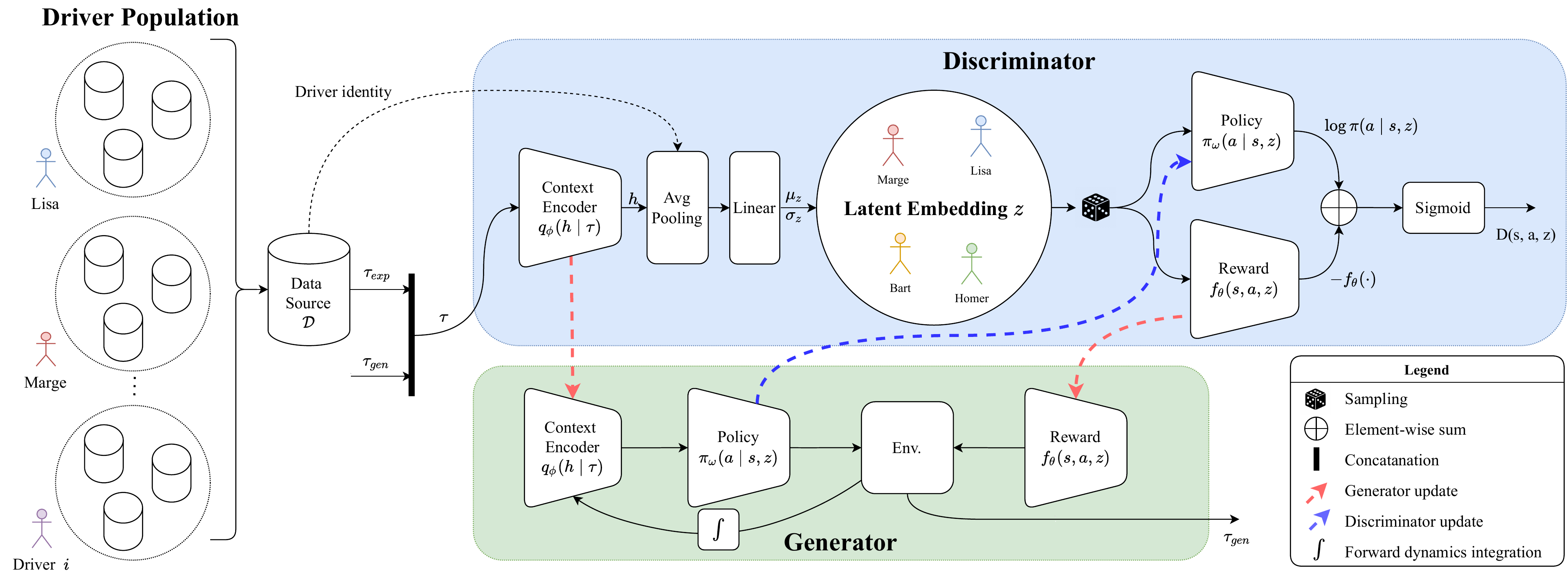}
    \caption{Our policy learning approach follows a latent form of adversarial IRL, where, at each training round, training alternates between a \textit{discriminator} to train a \textit{context encoder} (parameterized by $\phi$) and a \textit{reward} (parameterized by $\theta$), and a \textit{generator}, to train a \textit{policy} (parameterized by $\omega$). Prior to training, sequences of driving examples are stored for individual drivers in a population.  These examples are fed into a batch and fused with generated examples.  For those examples that are pooled together, the hidden context encoder states are passed to an average pooling layer, which averages those data labeled by a specific driver (denoted by the dotted black arrow), and whose output is the latent embedding for the individual drivers.  We subsequently use this embedding to determine a person-specific interaction policy, as discussed in Sec.~\ref{sec:intervention_policy}.
    }
    \label{fig:schema}
\end{figure*}

\section{RELATED WORK}
Previous work on devising human-understandable moves in games of Chess was introduced in~\cite{McIlroy-Young2020-kb}.  Many other papers have explored cooperative teaming using Theory-of-Mind to understand the human mental condition, in terms of benchmark frameworks~\cite{Shu2021-hm}, models for exploration and planning~\cite{Tsividis2021-rf}, and the watch-and-help challenge~\cite{Puig2020-sa}.  Some of these works have approached this from the perspective of an embodied AI system cooperating with humans~\cite{Carroll2020-oc}, while others have extended this basic premise to learn how to teach humans in cooperative tasks~\cite{Omidshafiei2018-lo}, while yet others have appealed to more data-free approaches~\cite{Strouse2021-aj}.

Approaches for learning personalized models of humans in an online fashion are also prevalent.  These include personality modeling for dialog systems~\cite{Yang2021-rj}, latent belief space planning under uncertainty~\cite{Qiu2020-ma}, and using latent factors in defining winning strategies for the AI system in zero-sum games with a human~\cite{Xie2020-dc}.  In driving and other motion prediction contexts, there are several examples of preference learning, including works that treat the problem as one of intent prediction~\cite{Lefevre2015-wi}, those involving Theory-of-Mind models mapped to a graph representation~\cite{Chandra2020-qa}, and those that involve few-shot learning~\cite{gui2018few}.
Several efforts have been made to characterize preference learning within reinforcement learning \cite{Akrour2012-hy,Wirth2017-bc}, including active learning approaches \cite{Sadigh2017-iz,Sadigh2016-zm} that are highly suited for human robot interactions, but do not translate easily to drivers on the road.

In the offline setting, several papers have addressed the problem of learning an individual among a population.  In the driving domain, the latent parameterization of a subject agent is learnt via an autoencoder in ~\cite{Morton2017-ob}. Other approaches focused on specific social states that pertain to right-of-way \cite{Schwarting2019} or maneuvers that are evident in relatively short horizons of behavior \cite{Deo2018-cm,Huang2020-pz}.

In our work, we aim to learn as much prior information as possible from observations of a population's behavioral history and track given drivers over several examples of an individual's driving behavior.  We focus on the problem of distinguishability of multiple possible factors, including those that tie directly to states that \textit{we should modify}. In particular, we learn whether an individual tends to become distracted which, in turn, gives insight into which intervention policy to use.

\section{BACKGROUND}
\label{sec:background}
\subsection{Adversarial Inverse Reinforcement Learning}
Our objective is based on the premise of learning a reward function from a dataset, which we cast as an inverse reinforcement learning (IRL) problem.  
We adopt an adversarial formulation of IRL~\cite{Fu2018-cc}, where the reward function is learned in a generative adversarial network (GAN).  This formulation is suited to data-rich applications and robust to recovering the actual reward, up to a constant.

\textbf{Markov Decision Processes.}
A Markov Decision Process (MDP) is defined by the tuple $\langle \calS, \calA, \calT, \calR, \gamma, \rho_0\rangle$, where $\calS$ is a state space, $\mathcal{A}$ is an action space, $\mathcal{T} : \calS \times \calA \times \calS \mapsto [0, 1]$ is a stochastic state transition function, $\mathcal{R} : \calS \times \calA \mapsto \mathbb{R}$ is a reward function, $\gamma \in [0, 1]$ is a discount factor, and $\rho_0(s) $ is an initial state distribution.  In the IRL setup, $\calR$ and $\calT$ are unknown, and must be recovered.

\textbf{MaxEnt IRL.} 
In maximum-entropy IRL, the objective is to find a policy $\pi^*$ that is exponentially close to the set of demonstration trajectories. In other words, it maximizes the entropy-regularized $\gamma$-discounted reward
\begin{equation} \label{eq:maxent}
\pi^* = \underset{\pi}{\arg\max\ } \mathbb{E}_\pi \left[\sum_{t=0}^T \gamma^t\calR(s_t, a_t) + \calH(\pi(a_t \mid s_t)) \right]
\end{equation}
where $\calH$ is the entropy of the policy distribution, i.e. $\calH(\pi) = \mathbb{E}_{\pi} \left[ -\log\pi \right]$ over a time horizon $T$.  The demonstrations are given by a set of trajectories, which denote the state-action pairs $\tau = (s_0, a_0, \ldots, s_T, a_T)$.  A collection of trajectories, complemented by an identifier of the human driver, $y$, forms a dataset of demonstrations $\calD$.  We assume that trajectories may be of different lengths.

In~\cite{Fu2018-cc}, the adversarial formulation of the MaxEnt problem is solved via a GAN.  Within the GAN, the \textit{discriminator} takes the form
\begin{equation}
D_\theta(s, a, s') = \frac{\exp(f_\theta(s, a, s'))}{\exp(f_\theta(s, a, s')) + \pi_\omega(a \mid s)},
\end{equation}
where $s' \sim \calT(s, a)$, $D_\theta$ is a discriminator network with parameters $\theta$, $f_\theta(\cdot)$ is factored as $f_{\theta_1, \theta_2}(s, a, s') = g_{\theta_1}(s, a) + \gamma h_{\theta_2}(s') - h_{\theta_2}(s)$, $\omega$ are the policy parameters, and $g_{\theta_1}$, $h_{\theta_2}$ are the base and potential components of the reward.

\textbf{Policy Optimization.} 
We use policy gradient methods to optimize $\pi$ in~\eqref{eq:maxent}, i.e. the \textit{generator} in the IRL problem.  One approach is proximal policy optimization (PPO)~\cite{schulman2017proximal}, where a policy update step applies a gradient of the objective $J$:
\begin{equation}
\nabla_\omega{J(\pi_\omega)} = \mathbb{E}_\omega \left[ \sum_{t=0}^{T-1} \gamma^t f_\theta(s_t, a_t, s_{t+1}) \nabla_\omega \log \pi_\omega(a_t \mid s_t) \right]
\end{equation}
which induces a gradient-ascent problem.


\section{PROBLEM STATEMENT}
\label{sec:problem}

We acquire a set of driving data $\calD$ from a variety of human drivers in various driving scenarios.  This data consists of the driver's actions, the motion of the vehicle, and interactions of a human driver with a human-machine interface (HMI) encoded as an HMI input.
Our goal is to learn a latent variable of \textit{driver traits}, $z$, that can help to govern how an HMI may best help a human driver, and an \textit{interaction policy}, $\pi(a^A|a^H, s, z)$, conditioned on knowledge of $z$ and an action of the human $a^H$ in some state $s$, where $a^A$ is an interaction choice.  We consider the simplest case where the action of the AI system $a^A$ is a binary signal passed to the driver, but this can be extended to any continuous or discrete signal to represent a wider range of interactions.

\textbf{Running example: A tale of four drivers.} 
We outline four types of drivers, each of who behave and respond differently to interactions to speed up or slow down during a highway merge scenario: 
\begin{enumerate}
    \item Lisa: a confident, often attentive driver who most likely ignores any HMI assistance.
    \item Marge: a cautious, often attentive driver, very receptive to HMI assistance.
    \item Bart: a confident, often distracted driver who most likely ignores HMI assistance.
    \item Homer: a cautious, often distracted driver, very receptive to HMI assistance.
\end{enumerate}
By learning to differentiate between these differing types of behavior and preference, we should be able to generate a more effective interaction policy.

\begin{figure*}
  \centering
  \includegraphics[width=0.75\textwidth]{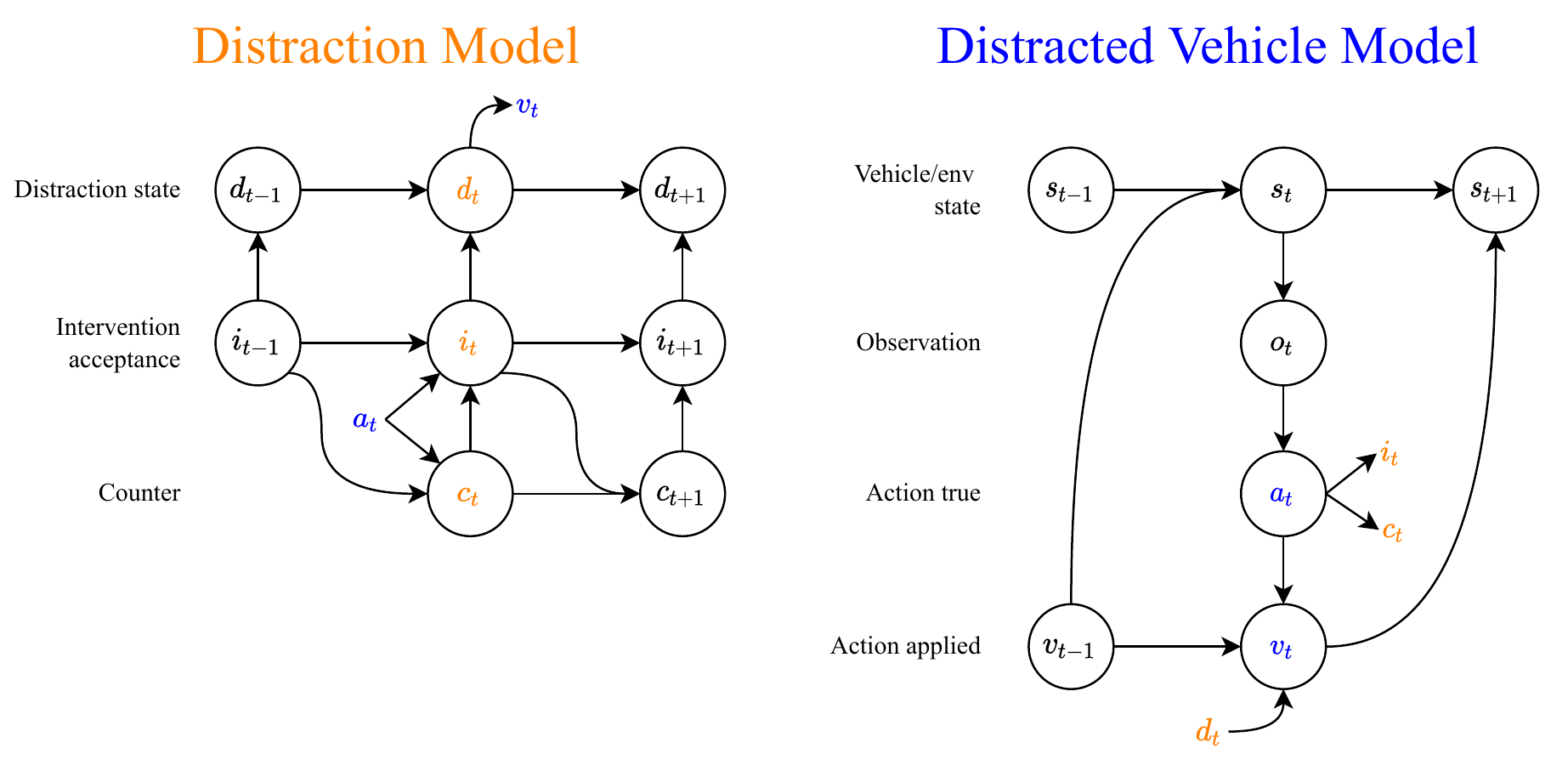}
  \caption{\textbf{Our implementation of a model of distracted driving.} Our model consists of two transition systems: a distraction model (left), which governs the tendency of a driver to become distracted and comply with or ignore an intervention, and a driving model (right) that models the vehicle's actions based on both the physical environment and distraction state.}
  \label{fig:distraction}
\end{figure*}

\section{LATENT MODELING OF COGNITIVE TRAITS}
\label{sec:approach}
Our insight is that modeling human driving behaviors can be treated as a meta-learning problem that imparts structure onto a latent representation according to the differences in behavior between individuals.
In general, the unique characteristics of individual drivers can easily be obscured by noise when observing a single time point, but when aggregating observations over many samples of driving, it becomes possible to distinguish one individual from another. This is especially true in the driving domain, where aggregation of several examples of one human's driving can illuminate individual characteristics of their behavior.  
We learn this structure using a dataset encompassing a population of drivers, and then exploit this structure to learn, and apply, the correct intervention policy. 
Although we do not explore it here, our learned human reward structures and policies are known to admit rich human-AI teaming and shared autonomy strategies~\cite{schaff2020residual, backman2022reinforcement}. 

\subsection{Trait- and Preference-Aware Adversarial IRL}
We adopt a meta-IRL approach~\cite{NEURIPS2019_30de2428} allowing to learn a reward structure and policy for the human from data, conditioned on the characteristics of particular humans.  The overall framework is depicted in Fig.~\ref{fig:schema}.  A latent variable $z$ is introduced that captures the differences in behaviors between humans, i.e. their traits.  Batches of time-series data of the human and their observations, each pre-identified per human, are fed to the learning framework.  During each epoch, training alternates between a discriminator and a generator.  The discriminator combines a \textit{context encoder} $q_{\psi}(h \mid \tau)$ that captures the traits of a driver and a \textit{reward function} $f_\theta(s,a,z)$ that encodes the rewards, penalties and preferences of the human agent.  Roll-outs generated from a frozen-parameter policy are used to compute the loss terms that allow for the joint optimization of context encoder and reward function. Upon convergence, the discriminator learns to distinguish between the driver-generated and policy-generated trajectories.
The discriminator takes the adversarial IRL form:
\begin{equation}
    D_\theta(s, a, s', z) = \frac{\exp(f_\theta(s, a, s', z))}{\exp(f_\theta(s, a, s', z)) + \pi_\omega(a \mid s, z)}    
\end{equation}

The generator, on the other hand, uses the frozen context encoder and reward function to train a \textit{trait-conditioned policy} $\pi_\omega(\cdot \mid s, z)$ for the human.  We describe each of these components below.



\textbf{Context encoder.}
The context encoder is represented as a network, $q_{\psi}(z \mid \tau)$, defining the probability of $z$ given the trajectory $\tau$.
We represent $q_{\psi}(z \mid \tau)$ as a long short-term memory (LSTM) network~\cite{hochreiter1997long}, taking in a trajectory history.  The hidden layer $h$ is fed into two linear layers representing the mean and log-variance of the latent encoding.  For all of our experiments, we adopt a single-hidden-layer LSTM with 128 units.

\textbf{Pooling layer.}
To account for sparsity of behavior indicative of traits, we pool together examples of drivers, determined based on a unique identifier such that we capture the \textit{aggregate} behavior of the human's history of behaviors within the latent embedding.  We insert this pooling layer just prior to the linear operations mapping LSTM hidden state $h$ to the statistics of $z$, i.e. $\mu_z$, $\log(\sigma_z)$, and hence it is subsumed within $q_\psi(\cdot)$.  Although we may use any aggregation operation, we choose an average-pooling operation over $h$ for our experiments.  We ensure that the pools are consistently sized by maintaining a fixed pooling size, and shuffling the batches for a particular human at each training round.

\textbf{Reward function.}
We adopt a discounted reward which maps traits, $z$, and states, actions, and next-states to a real-valued reward.  To remove reward ambiguity while learning, we adopt the form $f_{\theta_1, \theta_2}(s, a, s', z) = g_{\theta_1}(s, a, z) + \gamma h_{\theta_2}(s', z) - h_{\theta_2}(s, z)$~\cite{Fu2018-cc}.  We use a two-layer multi-layer perceptron (MLP) with 32 units per layer for our base reward network $g_{\theta_1}(\cdot)$ and a single-layer MLP with 32 units as our potential network $h_{\theta_2}(\cdot)$.

\textbf{Policy.}
Our policy is a 32-unit feed-forward network, $\pi_\omega(a \mid s, z)$, that encodes the probability of the human taking action $a$ given its perception of the world and their trait. 
We use the policy gradient algorithm proximal policy optimization (PPO)~\cite{schulman2017proximal} to train such a policy using the empirical reward structure and context encoder.

\textbf{Discriminator losses.}
Finally, we describe the losses used for training the discriminator.  See~\reftab{losses} for a complete list.  Loss $\calL_1$ encourages the likelihood of the trajectories under the reward, $p_\theta(\cdot)$ to match those of the demonstrated trajectories, $p_{\pi_h^*}(\cdot)$. Loss $\calL_2$ encourages high mutual information, $I_{p_\theta}$, between the trajectories and latent variables.
The contrastive loss $\calL_3$~\cite{HadsellCVPR2006} promotes cohesion of pairs of latent variables $z$, $z'$ having the same label ($y_z = y_{z'}$) and separation of pairs $z$, $z''$ with dissimilar labels ($y_z \neq y_{z''})$.  Here, labels can represent different driver types, different preferences for receiving AI help, and $\ell(\cdot, \cdot)$ may be any distance measure; we choose the $L_2$ norm.  One can soften the indicator function $\mathds{1}(\cdot)$ in order to more generally capture a continuum or partial ordering of labels. 
Finally, $\calL_4$ regularizes the latent variables according to a unit Gaussian.  We apply weights on the final three terms: 5.0 for $\calL_2$, 10.0 for $\calL_3$ and $10^{-4}$ for $\calL_4$.

\begin{table*}[t]
\centering
\caption{Discriminator losses.}
\label{tab:losses}
\begin{tabular}{ c l l }
\toprule
 $\calL_1 =$ & $\mathbb{E}_{p(z)}[D_{KL}(p_\theta(\tau \mid z) ~\Vert~ p_{\pi_h}(\tau \mid z))]$ & Reconstruct trajectory \\ \midrule
 $\calL_2 =$ & $-I_{p_\theta}(z, \tau)$ & M.I. between $z$ and trajectory $\tau$ \\    \midrule
  $\calL_3 =$ & $\sum_{z_0, z_1}\mathds{1}(y_{z_0} = y_{z_1})\ell(z_0, z_1)^2 + \mathds{1}(y_{z_0} \neq y_{z_1})\max(0, \epsilon - \ell(z_0, z_1))^2$ &  Contrastive loss between driver types or preferences \\\midrule
 $\calL_4 =$ & $\mathbb{E}_{p_\theta(\tau)}[D_{KL}(p_\theta(z \mid \tau) ~\Vert~ q_\phi(z \mid \tau))]$ & Latent regularization \\
 \bottomrule
\end{tabular}
\end{table*}

\section{MODELING A DISTRACTED DRIVING INTERVENTION SYSTEM}
\label{sec:intervention_policy}
We introduce a joint human-AI driving environment model, dubbed \texttt{HMIway-env}, built on \texttt{highway-env}~\cite{highway-env} to capture: (a) the cautiousness exhibited by a driver, (b) the likelihood of the driver to become distracted, and (c) the receptivity of a driver to an external alert issued by a vehicle AI system.  Our model captures momentary lapses in driving, such as when a driver is engaged in a secondary task (e.g. cellphone use) or is otherwise inattentive to the road.  It can also encompass delays in action that might result, for example, from the driver being engaged in conversation or in a state of high cognitive load~\cite{CHAN2013147}.

\texttt{Highway-env} is itself built on the OpenAI Gym environment~\cite{openaigym}, and is designed primarily for studying and training models for tactical decision-making tasks for autonomous driving.
The behavior of other vehicles on the road are modelled using the intelligent driver model~\cite{treiber2000} and initial speeds are randomized in a range within the prescribed speed limits of the road on which they are spawned.
Similar to the approach taken by Morton et. al.~\cite{Morton2017-ob}, the observations fed into the policy encode the positions and velocities of nearby vehicles as lidar observations. 



\textbf{Driving behavioral traits.}
We simulate three traits of interest, \textit{cautiousness}---the tendency of a human to treat the road scene with more or less caution, \textit{distractability}---the tendency of a human to become distracted while driving, and \textit{preference for AI help}.  To model cautiousness, we dilate the spatial footprint of the surrounding vehicles so that the perceived distance to them is reduced.  Accordingly, the \textit{obstacle inflation factor} is a multiplier placed on all objects to achieve this dilation.  This is meant to model the tendency of drivers to leave large headway gap with other drivers.  We make the assumption that cautiousness of a driver is directly linked to their tendency to prefer help from an assistive AI system.  We further model distractability via a controlled Markov model. In our model, see Fig.~\ref{fig:distraction}, this is represented by two states: a distraction state, $d_t$, encoding whether the driver is distracted or attentive, and an acceptance state, $i_t$ that encodes whether the driver has been influenced by the vehicle AI's alert signal.

\subsection{AI-Mediated Driver Distraction Model}\label{ssec:ai-mediated}
In Figure~\ref{fig:distraction} (right), the controlled Markov model consists of two actions: the vehicle action $a_t^{H}$, and the AI's intervention action $a_t^{A} \in \{alert, no\_alert\}$. We introduce a counter variable $c_t$ to encode how long a \textit{successful intervention by the vehicle AI} remains in effect. That is, when the vehicle AI issues an alert, the effect of the alert persists for a fixed time window, 
during which the acceptance state remains 1. 

In our model, the transitions for the acceptance state $i_t$ and $c_t$ are determined as follows:
\begin{flalign}
    \text{if} &\quad i_{t-1} = 0 \quad \text{then} \quad 
    \begin{cases}
          i_t = 1, c_t = 0 \quad &\text{if} \;  a_t^{A} = alert \\
          i_t = 0, c_t = 0 \quad &\text{if} \; a_t^{A} = no\_alert \\
    \end{cases}
    \label{eq:inter_1}
    \\
    \text{if} &\quad i_{t-1}=1, \quad a_t^{A} = alert, \quad \text{then} \quad c_t = 0 \quad \text{and} \quad i_t =1
    \label{eq:inter_2}
    \\
    \text{if} & \quad i_{t-1}=1, \quad a_t^{A} = no\_alert, \nonumber \\ 
    & \text{then} \quad c_t = (c_{t-1} + 1)\text{mod}\;N,   
    \begin{cases}
          i_t = 1 \quad &\text{if} \; c_{t-1} < N-1 \\
          i_t = 0 \quad &\text{if} \; c_{t-1} = N-1
    \end{cases}
    \label{eq:inter_3}
\end{flalign}

Relations~\eqref{eq:inter_1} and~\eqref{eq:inter_2} capture the behavior that, when a driver complies with an alert from the vehicle AI, their acceptance state is always set to be 1, regardless of the acceptance state at the previous time step. Additionally, $c_t$ is set to be 0, indicating that the vehicle AI's intervention has been re-triggered. If the driver's acceptance state is 0, then it continues to be 0 if no alert has been received. In~\eqref{eq:inter_3}, if the acceptance state is already 1 (which implies that the vehicle AI's alert was already accepted by the driver at an earlier time step), then the acceptance state continues to be 1 (that is the alert continues to have an effect on the driver) as long as $c_t$ remains within the \textit{intervention effectiveness window}. Once $c_t$ is greater than the window size, the acceptance state is reset to 0, if no more alerts are accepted.   

\begin{table}[t]
\caption{\label{tab:rewards}Driving related rewards for the joint human-vehicle AI system.}
\scriptsize{
\begin{tabular}{ll}
\toprule
$R_{\text{coll}}$ & $-5.0$ if crashed,  else 0 
\\
$R_{\text{speed}}$ & $5.0 * \frac{\textit{current\_speed}}{\textit{max\_speed}}$
\\
$R_{\text{right-lane}}$ & $0.1$, if in right lane, else 0
\\
$R_{\text{merging}}$ & $-0.1 * \frac{\textit{target\_speed} - \textit{current\_speed}}{\textit{target\_speed}}$ if in merging lane, else 0
\\
$R_{\text{lane-change}}$ & $-0.1$ if $a_t^{H}$ is $\textit{move\_left}$ or $\textit{move\_right}$, else 0
\\
$R_{\text{distraction}}$ & $-10.0$ if $d_t = 1$, else 0
\\
$R_{\text{alert}}$ & $10.0$ if $a_t^{A} = \textit{no\_alert}$ and $d_{t-1} = 0$, else 0\\
$R_{\text{accept-alert}}$ & $30.0$ if $a_t^{A} = \textit{alert}$ with $d_{t-1} = 1$ and $d_{t} = 0$, else 0\\ \bottomrule
\end{tabular}
}
\end{table}

\begin{table}[t]
\centering
\caption{Driver parameters for dataset generation and assistive policy learning.}
\label{tab:driver_params}
\begin{tabular}{lllll}
\toprule
Driver (ID)                                       & $\beta$ & $\alpha$ & $\eta$ & Infl. \\ 
\midrule
\textbf{Lisa (0):} $\downarrow$ distractibility, $\downarrow$ cautiousness    & 0.2      & 0.8     & 0.01     & 3.0          \\
\textbf{Marge (1):} $\downarrow$ distractibility, $\uparrow$ cautiousness      & 0.2      & 0.8     & 1.0      & 9.0          \\
\textbf{Bart (2):} $\uparrow$ distractibility, $\downarrow$ cautiousness & 0.8      & 0.2     & 0.01     & 3.0          \\
\textbf{Homer (3):} $\uparrow$ distractibility, $\uparrow$ cautiousness    & 0.8      & 0.2     & 1.0      & 9.0          \\ 
\bottomrule
\end{tabular}
\end{table}

\textbf{Mediating distraction.}
The distraction variable $d_t$ evolves as a controlled Markov chain whose transition probabilities are modulated according to the \textit{acceptance state} $i_t$ of the driver. The amount of modulation is controlled by the \textit{intervention effective factor}, $\eta$, which captures the driver's willingness to be influenced by the alert issued by the vehicle AI. 
Crucially, we assume that $\eta$ depends on the level of cautiousness of the driver. 
Upon accepting the alert from the vehicle AI, the baseline transition probabilities of the distraction state Markov chain are modulated in such a way the probability of becoming distracted is \textit{reduced} and of becoming more attentive is \textit{increased}. Specifically, if $\beta \in [0, 1]$ is the baseline probability of transitioning from an attentive state ($d_{t-1} = 0$) to a distracted state ($d_t = 1$), the conditional (modified according to the acceptance state) transition probabilities for the Markov model is given by
\begin{equation}\label{eq:dist_1}
    p(d_t = 1 | d_{t-1} = 0) = \textnormal{max}(0, \beta - \eta\mathds{1}(i_t = 1)).
\end{equation}
Likewise, the probability of transitioning from a distracted to an attentive state becomes higher when the driver is willing to accept the AI agent's alert. Therefore, 
\begin{equation}\label{eq:dist_2}
    p(d_t = 0 | d_{t-1} = 1) = \textnormal{min}(1, \alpha + \eta\mathds{1}(i_t = 1))
\end{equation}
where $\alpha \in [0, 1]$ is the baseline probability of transitioning from a distracted to an attentive state. 

From the above equations, we can see that if the driver accepts an alert from the vehicle's AI, the acceptance state will be set to 1, and as a result the transition probabilities in~\eqref{eq:dist_1} and~\eqref{eq:dist_2} are modulated and this modulation remains in effect for at least $N$ time steps. 

In our framework, at every timestep $t$, first $c_t$ is updated, followed by the acceptance state $i_t$ and then finally the distraction variable $d_t$. 

For $t > 0$, the value of $d_t$, the distraction state, affects the applied vehicle action $v_t$ as follows,
\begin{equation}
    v_t = \begin{cases}
          v_{t-1}\quad &\text{if} \, d_t = 1 \\
          a_t^{H} \quad &\text{if} \, d_t = 0
      \end{cases}
\end{equation}
with $v_0 = a_0^{H}$.

\begin{figure*}
    \centering
    \begin{subfigure}[b]{0.32\textwidth}
        \includegraphics[width=\textwidth, trim=30 0 60 0, clip]{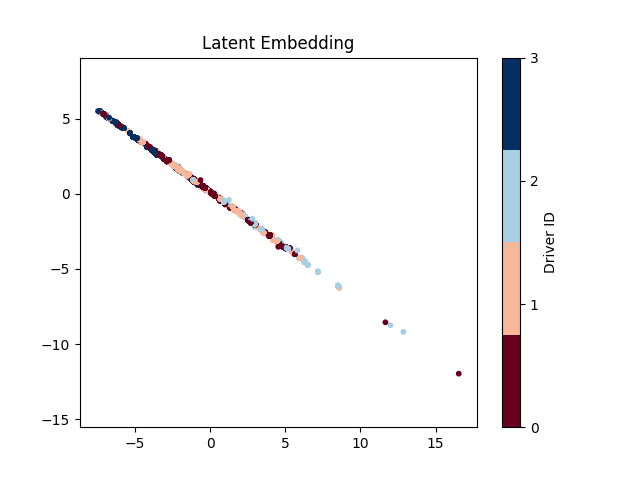}
        \caption{Unsupervised.}
    \end{subfigure}
    \begin{subfigure}[b]{0.32\textwidth}
        \includegraphics[width=\textwidth, trim=30 0 60 0, clip]{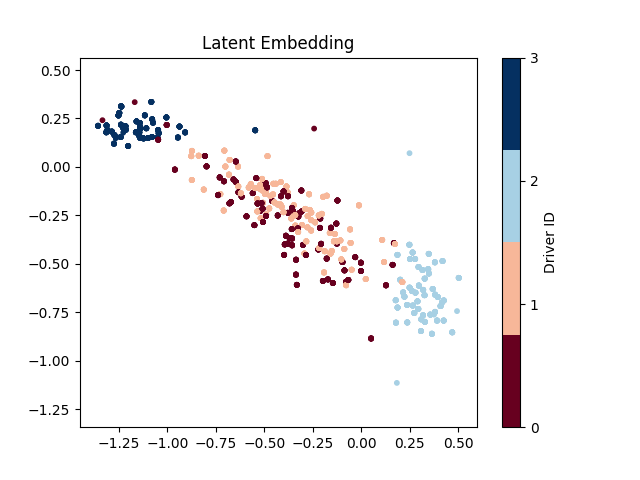}
        \caption{Supervision on ground-truth driver type.}
    \end{subfigure}
    \begin{subfigure}[b]{0.32\textwidth}
        \includegraphics[width=\textwidth, trim=30 0 60 0, clip]{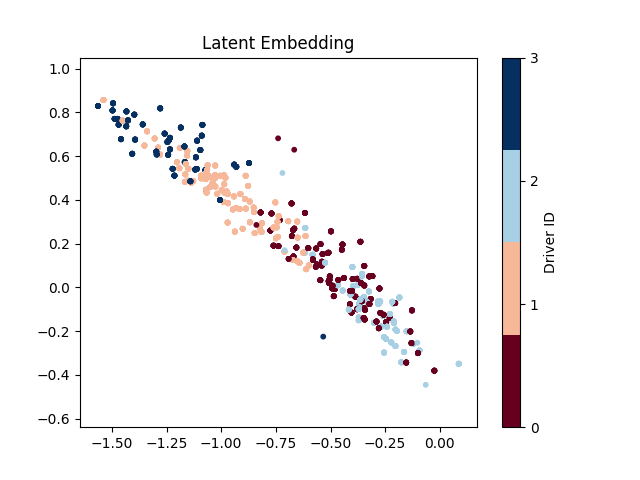}
        \caption{Supervision on ground-truth preferences.}
    \end{subfigure}
    \caption{Latent embeddings $z \in \mathbb{R}^2$ for three supervision experiments.  We illustrate the progression from unsupervised to supervised, demonstrating that supervision via either via labels or preferences is sufficient to distinguish different driver types.  The KL Divergence averaged between pairwise driver types was 435, 4940, and 2872, from (a) to (c).}
    \label{fig:latent_embeddings}
\end{figure*}

\textbf{Human subject validation.}
We lastly validated the model using a human studies experiment conducted on 500 individuals via a crowd-sourced survey.  Each individual was shown a small set of example roll-outs drawn from the model, and was asked to rate how safe, distracted, risky, or similar to their own driving the behavior was.  The findings of this study revealed that the intended behavior of the model was broadly consistent with human expectation of distracted and cautious driving, although further analysis is needed to account for judgement variations over a wider set of roll-outs.

\section{LEARNING AN INTERVENTION POLICY}
We frame the intervention as a reinforcement learning (RL) problem, and train a composite policy $\pi_H \circ \pi_{AI}$ of the human and AI system operating together.  Since we express both the human and AI system in terms of their higher-level goals and constraints in terms of a system of rewards, we may take the agent (the policy) to represent both the human and AI system operating together or two policies that interact with one another in a sequential manner. To generate our results, we opt for a 2-stage training protocol.

In the first stage, we train a driver policy that takes into account the different driver types. In the second stage, the HMI policy is trained on an environment that subsumes the driver policy. 
The combined human-AI policy consists of two action components: (a) the human-initiated vehicle actions ($a_{t}^{H}$) and  (b) the AI system's interventional actions on the driver ($a_t^{A}$). For human-initiated vehicle actions, the action space consists of a discrete set of semantic actions comprising \{\textit{speed\_up, slow\_down, keep\_speed, move\_left, move\_right}\}, where the first two actions result in changes in speed, while the last two bring about lane changes. The action space for the vehicle's interventions is binary and consists of a discrete set of interventional actions given by \{\textit{alert, no\_alert}\}. The mechanism by which $a_t^{A}$ affects the vehicle is facilitated by the intermediate distraction and alert-acceptance model described in Section~\ref{ssec:ai-mediated}.

\newcommand{\mytabhighspeedreturn}{
    \begin{tabular}{c|cccc}
    
        &   Lisa   &   Marge     &   Bart     &       Homer      \\
        \midrule
 Lisa  &   445      &   *       &       *       &       * \\
 Marge    &   *       &   446     &       *       &       * \\
 Bart    &   *       &   *       &       444      &       * \\
 Homer    &   *       &   *       &       *       &      458 \\
 AvgHMI    &   447       &   421      &       432      &      391 \\
 NoHMI      &  451       &   423       &       428       &       404 \\
\bottomrule
\end{tabular}
}

\newcommand{\mytabdistractionreward}{
    \begin{tabular}{c|cccc}

        &   Lisa   &   Marge     &   Bart     &       Homer      \\
        \midrule
 Lisa  &   -85      &   *       &       *       &       * \\
 Marge    &   *       &   -80     &       *       &       * \\
 Bart    &   *       &   *       &       -123      &       * \\
 Homer    &   *       &   *       &       *       &       -117 \\
 AvgHMI    &   -181       &   -45      &       -744      &       -162 \\
 NoHMI      &   -192       &   -176       &       -745       &       -770 \\
\bottomrule
\end{tabular}
}

\begin{table*}[t]%
  \centering
  \caption{Metrics for human-AI models trained in the merge environment for the four different driver types. The driver parameters are same as in Table~\ref{tab:driver_params}. (Left) Average high speed return (Right) Average total distraction reward per episode. Rows 1-4 are driver-specific HMI models. Row 5 is an HMI model tuned to an `average' driver type with (Infl. = 6.0, $\eta$ = 0.505, $\beta = 0.5$, $\alpha = 0.5$) and Row 6 is with no HMI assistance.} %
    \mytabhighspeedreturn
    \;
    \mytabdistractionreward
  \label{tbl:metrics_table}%
\end{table*}

\subsection{Reward Structure}
Table~\ref{tab:rewards} shows the full list of driving-related rewards used by the joint human-vehicle AI model capturing a joint policy trained for the actions of both the human and AI system. 

$R_{\text{coll}}$, $R_{\text{speed}}$, $R_{\text{right-lane}}$, $R_{\text{merging}}$, $R_{\text{lane-change}}$ are the reward components pertaining to driving performance, while $R_{\text{distraction}}$ pertains to the human and $R_{\text{alert}}$ captures the rewards the vehicle AI receives for issuing sparse alerts. $R_{\text{accept-alert}}$ is the reward term that connects the vehicle AI to the human. 

For training our models, we set the coefficients related to driving rewards to values such that the vehicle favors being safe on the right lane, at higher speeds and seeks to minimize lane changes.

\section{EXPERIMENTS}
\label{sec:experiments}


\subsection{Dataset Generation and Latent Trait Learning}
To generate training data, we sample from four driver types spanning levels of distraction and levels of cautiousness, as per our running example.  We generate $3\times10^5$ timesteps per type, according to the values in Table~\ref{tab:driver_params}.  We generate episodes based on a merge scenario, pictured in Fig.~\ref{fig:teaser}, having three available lanes, and with a maximum of 20 randomly-placed vehicles.  The actions are stepped at a rate of 5 Hz, while the simulator is updated at 15 Hz.

We consider this dataset as ground truth, and execute training based on three assumptions on what aspects of the data are labeled.
\begin{description}[style=unboxed]
    \item[Unsupervised.]  We assume no supervision on the data, and allow the latent embedding to be learned purely based on the learner's ability to detect nuances in behavior.
    \item[Supervised on driver ID.] We train in the general case where the types of drivers are fully labeled.  
    \item[Supervised on preferences.] We assume preferences of the driver are identifiable.  We argue that such labels could be obtained more easily than for the case with ID supervision, via preference queries given to a subset of drivers~\cite{Sadigh2017-iz}.
\end{description}

We build training data for each case and assume that 20\% of the data is labeled.  We use two discriminator updates per training round with a learning rate of $5\times 10^{-4}$, and allow the generator updates for 500 steps with a learning rate of $10^{-2}$.  The LSTM takes in a trajectory 20 seconds in length, and we use a pooling size of 8 for each driver ID.  For purposes of comparison, we terminate training after $10^6$ timesteps.  In all experiments, we use a two-dimensional latent space.

\subsection{Distinguishability}
In Fig.~\ref{fig:latent_embeddings}, we notice that the ability to distinguish driver types is heavily influenced by the level of supervision.  In the case of driver ID supervision, our training was successful at achieving good separation of latent variables associated with high-distraction levels from those that were lower (comparing cases 0 (Lisa) and 1 (Marge) with 2 (Bart) and 3 (Homer)).  The two high-distraction cases (2 and 3) are also highly separated.  In the case of ground-truth preferences, each of the cases yield latent embeddings that are relatively-well clustered; the 
data
indicate that the two distinct preference settings separate into clusters, with more separation happening when the driver tends to be more distracted.  We hypothesize that the failure to distinguish between different confidence levels in low-distraction cases is likely due to the chosen scenario.

\subsection{Impact of HMI Interventions}

Table~\ref{tbl:metrics_table} presents the performance of (1) driver-specific HMI models (rows 1-4) (2) an HMI tuned to an `average` driver type (Infl. = 6.0, $\eta$ = 0.505, $\beta = 0.5$, $\alpha = 0.5$), denoted as \textit{AvgHMI} assisting the four driver types (row 5) and (3) the drivers with no HMI assistance (row 6). The performance is evaluated by computing the average high speed return (Table~\ref{tbl:metrics_table}, left) and the average distraction reward per episode (Table~\ref{tbl:metrics_table}, right).

We observe that the distraction rewards are the highest when the drivers are assisted by driver-specific HMI models.  In particular, we find that the average total distraction reward over an episode is -101.2 for personalized HMI, compared with -283.0 for \textit{AvgHMI} and -470.8 for \textit{NoHMI}. Additionally, Bart (less cautious, less receptive and highly distracted) becomes distracted the most across all drivers regardless of the HMI model used. We also see that when the \textit{AvgHMI} assists Marge and Homer, who are both highly receptive to HMI assistance, the average model is able to do a good job in keeping the distraction rewards fairly high. Since the \textit{AvgHMI} is trained to assist an `average' driver who tends be more distracted than Marge, but less than Homer, it is effective in increasing Marge's distraction reward even better than the Marge-specific HMI model. However we do see that this comes at a cost: when \textit{AvgHMI} assists Marge and Homer, the high speed returns are lower than driver-specific models. The decrease is more prominent for Homer who is much more distractible. 

With regards to high speed returns, we observe that in general, driver-specific HMI models are able to help the drivers achieve consistently high speed returns. In particular, we see an average return of 448 over the four individuals, compared with 423 for the \textit{AvgHMI}, and 427 for \textit{NoHMI}. Except for Lisa, for all other drivers, there is a decrease in the high speed return when there is no assistance.

\section{CONCLUSIONS}
We explore a framework to capture and use latent representations of individual drivers to help decide the best strategy to assist them.  We furthermore develop a simulator environment that allows for data generation from a diverse set of driver types.  Our experiments show that we can uncover the latent structure of different drivers and show that the representation may be useful for interaction.  We further show that our intervention strategies produce higher rewards than an HMI system designed for an average-case human.

In our work, we assume explicit structure embedded in both the data generating model and the labels. Future work will include more implicit understanding of preferences from population data that includes interactions with an AI system, where we can explore the emerging structure of human behavior from such datasets, and the interplay of weak supervision and explainability.  
We also aim to leverage the rewards and policies learned from humans, in addition to their latent representations, to train intervention policies and pave the way toward more personalized interventions.


\addtolength{\textheight}{-12cm}   







\bibliographystyle{IEEEtran}
\bibliography{references}

\end{document}